\documentclass[%
 reprint,
 amsmath,amssymb,
 superscriptaddress,
 aps,
]{revtex4-2}

\usepackage{graphicx}
\usepackage{blindtext}
\usepackage{dcolumn}
\usepackage{bm}
\usepackage{braket}
\usepackage{setspace}
\usepackage{color}
\usepackage{mathtools}
\usepackage{float}

\usepackage[english]{babel}

\begin{document}
%\begin{spacing}{1.5}
\preprint{APS/123-QED}

\title{
Performance Benchmarking of Quantum Algorithms for Hard Combinatorial Optimization Problems: A Comparative Study of non-FTQC Approaches
}

\author{Santaro Kikuura}
\affiliation{
Department of Computer Science and Engineering, Shibaura Institute of Technology, Toyosu, Tokyo 135-8548, Japan
}

\author{Ryoya Igata}
\affiliation{
Department of Computer Science and Engineering, Shibaura Institute of Technology, Toyosu, Tokyo 135-8548, Japan
}

\author{Yuta Shingu}
\affiliation{
 Department of Physics, Tokyo University of Science,
 Shinjuku, Tokyo, 162-8601, Japan
}

\author{Shohei Watabe}
\affiliation{
Department of Computer Science and Engineering, Shibaura Institute of Technology, Toyosu, Tokyo 135-8548, Japan
}

\begin{abstract}
This study systematically benchmarks several non-fault-tolerant quantum computing  algorithms across four distinct optimization problems: max-cut, number partitioning, knapsack, and quantum spin glass. 
Our benchmark includes noisy intermediate-scale quantum (NISQ) algorithms, such as the variational quantum eigensolver, quantum approximate optimization algorithm, quantum imaginary time evolution, and imaginary time quantum annealing, 
with both ansatz-based and ansatz-free implementations, 
alongside tensor network methods and direct simulations of the imaginary-time Schr\"odinger equation. 
For comparative analysis, we also utilize classical simulated annealing and quantum annealing on D-Wave devices. 
Employing default configurations, our findings reveal that no single non-FTQC algorithm performs optimally across all problem types, underscoring the need for tailored algorithmic strategies. This work provides an objective performance baseline and serves as a critical reference point for advancing NISQ algorithms and quantum annealing platforms. 
\end{abstract}

\maketitle

%\tableofcontents

\section{Introduction}
The rapid advances in quantum computer science is transforming traditional computational paradigms and opening new possibilities across various fields, including quantum chemistry, materials science, cryptography, machine learning, and combinatorial optimization. A
mong them, combinatorial optimization poses significant challenges due to its NP-hard classification, a characteristic shared by many complex real-world problems in science and engineering. 
Quantum algorithms offer attractive alternatives by leveraging unique quantum properties, which may allow them to outperform classical approaches in this domain.

In particular, non-fault-tolerant quantum computer (non-FTQC), implemented on noisy intermediate-scale quantum (NISQ) devices~\cite{Preskill2018quantumcomputingin} and quantum annealing platforms like D-Wave devices, has emerged 
as a compelling approach for hopefully enhancing computational efficiency in optimization tasks. 
In particular, 
quantum algorithms designed specifically for non-FTQC devices, such as the quantum annealing (QA)~\cite{kadowaki1998quantum,RevModPhys.90.015002}, the variational quantum eigensolver (VQE)~\cite{Peruzzo2014}, the quantum approximate optimization algorithm (QAOA)~\cite{farhi2014QAOA}, and quantum imaginary-time evolution (QITE)~\cite{McArdle2019,Yuan2019theoryofvariational,Motta2020}, have gained significant attention. 
While each algorithm employs a distinct approach, VQE, QAOA, and QITE have been developed specifically with NISQ devices, making them suitable for near-term quantum applications.

While existing studies have benchmarked specific algorithms on individual optimization problems --- such as QA and QAOA for the max-cut problem~\cite{10.1145/3678184}, QA and simulated annealing (SA) for Ising models~\cite{doi:10.1126/science.220.4598.671,PhysRevA.96.022326}, and QA, SA, and the Gurobi optimizer~\cite{gurobi} for the cardinality-constrained quadratic knapsack problem~\cite{Ceselli2023} --- comprehensive comparative studies across a broader range of non-FTQC algorithms are still limited. Such analyses are particularly valuable in the context of combinatorial optimization problems as well as the quantum computer science. 

Additionally, imaginary-time-based methods have shown promise for global energy minimum searches~\cite{Amara1993}, with imaginary time quantum annealing (ITQA)~\cite{Nishimori_2008} demonstrating advantages over QA in Grover’s algorithm~\cite{10.1063/1.2995837}. 
Similarly, tensor network methods enable efficient simulations of quantum dynamics~\cite{10.21468/SciPostPhysCodeb.4}, though their application remains relatively underexplored compared to other non-FTQC approaches in combinatorial optimization contexts.

Given the diversity of optimization problems and the distinct requirements of different quantum algorithms, a comprehensive performance evaluation of these algorithms is crucial. In this study, we benchmark VQE, QAOA, QITE, ITQA, QA, and additional methods like SA and tensor networks across four representative optimization problems: max-cut, number partitioning, knapsack, and quantum spin glass. Through this systematic analysis, we aim to clarify the strengths and limitations of both NISQ-compatible and quantum annealing-based approaches, providing insights to guide the development of quantum algorithms for complex optimization tasks. We establish a controlled baseline that allows for an objective comparison of their efficiency and suitability across various problem types. Our findings suggest that no single non-FTQC algorithm excels across all problems, underscoring the need for adaptive strategies and targeted algorithmic designs. This baseline serves as a critical foundation for evaluating future advancements, facilitating a more complete understanding of the potential of non-FTQC algorithms.

\section{Benchmark Problem Set and Hamiltonians}
We employ four distinct problems for performance benchmarking, comprising three NP-complete problems~\cite{10.3389/fphy.2014.00005} and one quantum problem. In all cases, we aim to minimize a cost or energy function.

The first problem is the max-cut problem. 
Consider an undirected graph $G = (V, E)$ with vertex set $V$ and edge set $E$. 
The goal is to partition $V$ into two subsets such that the number of edges between these subsets is maximized. 
The problem Hamiltonian is formulated as
\begin{align}
    \hat{H}_{\mathrm{prob}} = - \frac{1}{2} \sum\limits_{(i,j) \in E} \left(1 - \hat{\sigma}_i^z \hat{\sigma}_j^z \right),
\end{align}
where $\hat{\sigma}_i^z$ denotes the Pauli-$z$ operator on vertex $i$.

The second problem is the number partitioning problem. 
Consider a set $S = (n_1, n_2, \dots, n_N)$ of $N$ positive numbers. 
The objective is to partition $S$ into two subsets such that the sums of their elements are as close as possible. 
The problem Hamiltonian is given by
\begin{align}
    \hat{H}_{\mathrm{prob}} = \left( \sum\limits_{i=1}^N n_i \hat{\sigma}_i^z \right)^2.
\end{align}

The third problem is the knapsack problem. 
Given a set of $N$ items, 
each of which are characterized by a weight $w_i$ and value $c_i$ for $i = 1, \dots, N$, 
the task is to select items to maximize the total value in the knapsack while ensuring the total weight does not exceed a specified capacity $W$. The Hamiltonian for this problem is expressed as
\begin{align}
    \hat{H}_{\mathrm{prob}} = - \sum\limits_{i=1}^N c_i \hat{q}_i + A \left( \sum\limits_{i=1}^N w_i \hat{q}_i - W \right)^2,
\end{align}
where $\hat{q}_i \equiv (1 + \hat{\sigma}_i^z)/2$. 
The first term represents the total value of items in the knapsack, 
while the second term is a penalty to enforce the weight constraint, with the penalty strength set as $A = 2  \max_i (c_i)$ in this study.

The final problem is the quantum spin glass model. 
The Hamiltonian for this problem is
\begin{align}
    \hat{H}_{\mathrm{prob}} = \sum\limits_{i,j} \left( J_{ij}^x \hat{\sigma}_i^x \hat{\sigma}_j^x + J_{ij}^y \hat{\sigma}_i^y \hat{\sigma}_j^y + J_{ij}^z \hat{\sigma}_i^z \hat{\sigma}_j^z \right),
\end{align}
where $\hat{\sigma}_i^{y,z}$ are Pauli-$y$ and Pauli-$z$ operators, respectively. 
In this study, couplings $J_{ij}^{\alpha}$ for $\alpha = x,y,z$ are randomly drawn from a truncated normal distribution with mean $\mu = 0$, standard deviation $\sigma = 0.3$, and the interval $(x_{\rm min}, x_{\rm max}) = (-1, 1)$.

These four problems were selected for their distinct structural characteristics. 
The max-cut problem is relatively sparse and not fully connected. 
The number partitioning problem is fully connected. 
The knapsack problem is also fully connected and includes a penalty term. 
In contrast to the previous three, the quantum spin glass problem is both fully connected and inherently quantum.

\section{Non-FTQC Algorithms for Ground State Search and Combinatorial Optimization}

Non-FTQC algorithms have been intensively applied to ground state search problems, making them especially valuable in combinatorial optimization, where classical methods often face significant limitations. 
Early approaches in this domain include quantum annealing, which leverages adiabatic evolution to find ground states of problem Hamiltonians~\cite{kadowaki1998quantum}. 
More recently, the QAOA~\cite{farhi2014QAOA} and the VQE~\cite{Peruzzo2014} have gained attention. 
Both NISQ algorithms optimize parameter sets to minimize the expectation value of a given problem Hamiltonian. 
Additionally, techniques based on imaginary time evolution~\cite{McArdle2019,Yuan2019theoryofvariational,Motta2020} expand the set of available quantum techniques for combinatorial optimization~\cite{bauer2023combinatorialoptimizationquantumimaginary} by utilizing non-unitary evolution to approximate ground states. 
In this section, we provide an overview of these algorithms, discussing their foundational principles and objective functions.

In the QAOA~\cite{farhi2014QAOA}, the objective function is the expectation value of the problem Hamiltonian $\hat{H}_{\rm prob}$ whose ground state encodes the solution, expressed as
$F_p (\boldsymbol{\gamma}, \boldsymbol{\beta}) = \langle \boldsymbol{\gamma}, \boldsymbol{\beta} |  \hat{H}_{\rm prob} | \boldsymbol{\gamma}, \boldsymbol{\beta} \rangle$,
where $\boldsymbol{\gamma}$ and $\boldsymbol{\beta}$ are parameter sets defined as $\boldsymbol{\gamma} = (\gamma_1, \dots, \gamma_p)$ and $\boldsymbol{\beta} = (\beta_1, \dots, \beta_p)$. The parameterized state \( | \boldsymbol{\gamma}, \boldsymbol{\beta} \rangle \) is constructed via
\begin{align}
    | \boldsymbol{\gamma}, \boldsymbol{\beta} \rangle 
    = 
    \hat{U} (\beta_p) \hat{U} (\gamma_p) \cdots \hat{U} (\beta_1) \hat{U} (\gamma_1) | +\rangle^{\otimes N},
\end{align}
with operators $\hat{U} (\beta_i) = e^{-i \beta_i \hat{H}_{\rm mix}}$ and $\hat{U} (\gamma_i) = e^{-i \gamma_i \hat{H}_{\rm prob}}$, where $| + \rangle^{\otimes N}$ denotes the equal superposition state. Here, $\hat{H}_{\rm mix} \equiv \sum_{i=1}^N \hat{\sigma}_i^x$ is the mixing Hamiltonian. The optimal parameters $(\boldsymbol{\gamma}^*, \boldsymbol{\beta}^*)$ are obtained by minimizing $F_p(\boldsymbol{\gamma}, \boldsymbol{\beta})$, i.e., 
$(\boldsymbol{\gamma}^*, \boldsymbol{\beta}^*) = {\rm arg } \, \min_{\boldsymbol{\gamma}, \boldsymbol{\beta}} F_p (\boldsymbol{\gamma}, \boldsymbol{\beta})$.

The VQE~\cite{Peruzzo2014}, another algorithm suited to near-term quantum devices, also seeks to minimize the expectation value of the problem Hamiltonian. Here, the goal is to find the optimal parameters
\begin{align}
    \boldsymbol{\theta}^*
    = 
    {\rm arg } \, \min_{\boldsymbol{\theta}} 
    \langle \Psi (\boldsymbol{\theta} ) | 
    \hat{H}_{\rm prob} | \Psi (\boldsymbol{\theta} ) \rangle,
\end{align}
where $ | \Psi (\boldsymbol{\theta}) \rangle = \hat{U} (\boldsymbol{\theta}) | + \rangle^{\otimes N}$  and the circuit operator $ \hat{U}(\boldsymbol{\theta})$ is parameterized by $\boldsymbol{\theta}$. 
Here, we initialize the state as $| + \rangle^{\otimes N}$, encoding a uniform distribution over solution states, advantageous in combinatorial optimization contexts.

For QITE on near-term devices, two main approaches have been proposed: ansatz-based QITE~\cite{McArdle2019,Yuan2019theoryofvariational} and ansatz-free QITE~\cite{Motta2020}. 
The ansatz-based QITE method leverages the imaginary-time Schr\"odinger equation,
\begin{align}
    \frac{d}{dt} | \Psi (t) \rangle  = - \hat{H} | \Psi (t) \rangle,
    \label{QITE}
\end{align}
which, due to its non-unitary nature, cannot be directly implemented with standard unitary gates. 
To circumvent this, we introduce a variational approach rooted in MacLachlan's variational principle~\cite{MacLachlan1964}, which minimizes the objective function as
\begin{align}
    \delta \left\| \left(\frac{d}{dt} + \hat{H} \right) | \Psi ( \boldsymbol{\theta}(t) ) \rangle \right\| = 0,
\end{align}
where \(| \Psi ( \boldsymbol{\theta}(t) ) \rangle\) represents the state generated by the ansatz circuit with time-dependent parameters \(\boldsymbol{\theta}(t)\). The optimization ensures that the evolution governed by \(\boldsymbol{\theta}(t)\) closely approximates the non-unitary imaginary time dynamics.

The ansatz-free QITE method~\cite{Motta2020} approximates the imaginary time evolution operator $e^{ - \beta \hat{H} }$ by applying a Trotter decomposition. After a single Trotter step, the state evolves to \( | \Psi' \rangle = e^{-\Delta \tau \hat{H}} | \Psi \rangle \), where \( | \Psi' \rangle \) represents the state after an imaginary time increment \( \Delta \tau \). 
This state, however, must be approximated using unitary evolution on quantum hardware. 
To achieve this, the imaginary time step is approximated by a unitary operator \( e^{-i \Delta \tau \hat{A}[m]} \), where \(\hat{A}[m]\) is expressed as a sum of Pauli strings:
$\hat{A}[m] = \sum_{i_1, \dots, i_k} a[m]_{i_1, \dots, i_k} \hat{\sigma}_{i_1} \dots \hat{\sigma}_{i_k}$.
Here, the coefficients \(a[m]_{i_1, \dots, i_k}\) are determined to minimize the distance between the two states 
\begin{align}
    \left\| | \Psi' \rangle - e^{-i \Delta \tau \hat{A}[m]}  | \Psi \rangle \right\|,
\end{align}
thus ensuring that the unitary evolution with \(\hat{A}[m]\) closely approximates the non-unitary imaginary time evolution.

The QA~\cite{kadowaki1998quantum} employs a time-dependent Hamiltonian given by
\begin{align}
    \hat{H}_{\rm QA}(t) = \frac{t}{T} \hat{H}_{\rm prob} + \left( 1 - \frac{t}{T} \right) \hat{H}_{\rm d},
    \label{QA_Hamiltonian}
\end{align}
where $\hat{H}_{\rm prob}$ represents the problem Hamiltonian, 
and $\hat{H}_{\rm d} = - \sum_{i=1}^{N} \hat{\sigma}_i^x$ is the driver Hamiltonian. 
The process begins with the initial state set as the ground state of $\hat{H}_{\rm d}$, which corresponds to the equal superposition state $ | + \rangle^{\otimes N} $. 
Through adiabatic evolution, as \(t\) progresses from 0 to \(T\), the system remains in the instantaneous ground state, ultimately reaching the ground state of $\hat{H}_{\rm prob}$ 
if the evolution is adiabatically slow enough. 
This process is governed by the real-time Schr\"odinger equation,
\begin{align}
    i \frac{d}{dt} | \Psi (t) \rangle = \hat{H}_{\rm QA}(t) | \Psi (t) \rangle.
\end{align}

The ITQA extends the conventional QA approach by incorporating imaginary time evolution to increase the ground state population more effectively. 
In ITQA, we solve the imaginary-time Schr\"odinger equation \eqref{QITE},
using the QA Hamiltonian~\eqref{QA_Hamiltonian}. 
Unlike the QITE protocol, which applies imaginary time evolution with a time-independent problem Hamiltonian, ITQA begins with the driver Hamiltonian with the equal-superposition ground state and adiabatically interpolates towards the problem Hamiltonian. 
This gradual shift enhances the ground state convergence, with leveraging the benefits of imaginary time dynamics to increase the population in the target ground state.

% benchmark 
\begin{figure*}[t]
    \centering
    \includegraphics[keepaspectratio, width=176.5mm]{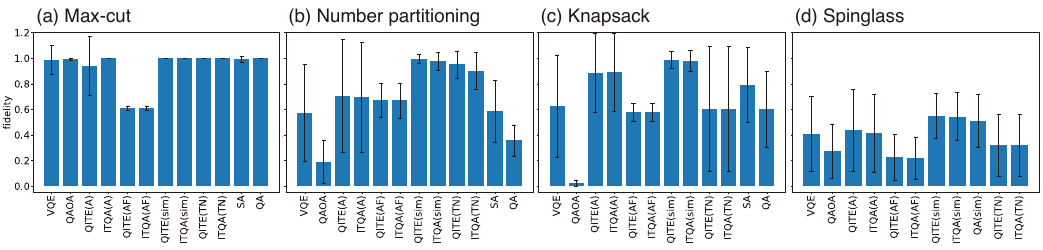}
\caption{
Performance benchmarking results for various optimization problems: (a) max-cut, (b) number partitioning, (c) knapsack, and (d) quantum spin glass. Benchmarked algorithms include VQE, QAOA, SA, QA, and both QITE and ITQA, evaluated under four configurations: ansatz-based (A), ansatz-free (AF), direct solution of the imaginary time Schr\"odinger equation (sim), and tensor network (TN) methods. Note that the quantum spin glass model cannot be addressed on D-Wave devices, so QA results are based on simulations in the panel (d).
}
    \label{fig:1}
\end{figure*}

\section{Performance Benchmarking of Non-FTQC Algorithms}

To evaluate the performance of non-FTQC algorithms, we benchmark the VQE, QAOA, ansatz-based QITE and ITQA, ansatz-free QITE and ITQA, as well as QA. 
For NISQ algorithms, we focus on their idealized performance using simulations performed with Qiskit~\cite{javadiabhari2024quantumcomputingqiskit}. Conversely, for QA, we utilized D-Wave's quantum devices. For further comparative analysis, we simulated imaginary-time evolution for QITE and ITQA by directly solving the imaginary-time Schr\"odinger equation using QuTiP~\cite{JOHANSSON20121760,JOHANSSON20131234}, as well as the tensor network approach with iTensor~\cite{10.21468/SciPostPhysCodeb.4,10.21468/SciPostPhysCodeb.4-r0.3}, and we also simulated conventional classical SA.

In the QAOA, we set the circuit depth as $p = 100$, employing the COBYLA (Constrained Optimization BY Linear Approximations) optimizer with a maximum of $5000$ iterations. For VQE, QITE, and ITQA under the ansatz-based approach, we utilized the efficient SU(2) circuit with two repetitions, applying a Hadamard gate to all qubits before introducing the efficient SU(2) gates to prepare equal superposition states, thus including solution states as in QA. For VQE, we adopted the Simultaneous Perturbation Stochastic Approximation (SPSA) optimizer, also with $5000$ maximum iterations. 
For QA, we used the Advantage2\_prototype1.1 QPU in D-Wave diveces, with an annealing time of $20 \, \mu \text{s}$. 
SA involved $10000$ sweeps from a random initial state, with $1000$ shots for both QA and SA. 
For tensor network simulations in iTensor, we set the bond dimension to the problem size $N$, matching the number of qubits $N$ across non-FTQC algorithms.
In performance benchmarking, we generated $250$ instances each for the max-cut, number partitioning, knapsack, and quantum spin glass problems. 
All the problems are with the size $N=5$, which is a feasible size for exhaustive testing across algorithms and problem types.

For QITE and ITQA, we determine the evolution time $T$ and the time-slice size $\Delta t$ to avoid disruptive oscillatory behavior in the energy expectation value during imaginary-time evolution. 
Due to drastic differences in energy scales across problem types (i.e., max-cut, number partitioning, knapsack, and quantum spin glass), we employed distinct $\Delta t$ and $T$ values for each problem types. 
We chose $\Delta t$ that ensures the smooth imaginary-time dependence in the energy expectation value across all $250$ instances, adjusting $T$ to show the sufficient convergence in the single instance. 
Here, we employ consistent time-steps across problem types. 
Once optimized for ansatz-based QITE and ITQA, these parameters $\Delta t$ and $T$ are applied consistently to ansatz-free algorithms and tensor network and QuTiP simulations.

A criterion of $\Delta t$ ensuring smooth time-evolution was found to be $|E| \Delta t \lesssim 10^{-1}$, where $E$ is the typical system energy such as the expectation value of the Hamiltonian. 
For example, in this study, we used $(\Delta t, T) = (10^{-1}, 10^3)$ for the max-cut problem 
and the quantum spin glass problem both with $|E|=O(1)$, 
while for the number partitioning problem with $|E| = O(10^4)$, 
we choose $(\Delta t, T) = (10^{-5}, 0.2)$. 
The knapsack problem ($|E| = O(10^6)$) follows with $(\Delta t, T) = (10^{-7}, 10^{-3})$. 
For consistency, we used $10^4$ time steps, except in number partitioning, which required double this amount due to relatively slow convergence at $T = 0.1$.

Figure~\ref{fig:1} summarizes the performance benchmarking results. Here, QA fidelity in D-Wave experiments and SA is defined as the fraction of instances yielding optimal solutions over $1000$ shots. The max-cut problem, sparse in structure, yielded high performance across most algorithms. The number partitioning and knapsack problems exhibited intermediate difficulty, while quantum spin glass models were challenging for nearly all algorithms.

For max-cut problem, 
ansatz-free algorithms exhibited lower fidelity compared to other methods, with notable variance in VQE and ansatz-based QITE results. 
Here, while the energy expectation was close to the ground state, certain instances were trapped in an excited state. 
The variance seen in some other problem types follows a similar pattern, where energy is near ground state levels, yet the state becomes stuck orthogonal to the true ground state. 
Direct imaginary-time Schrödinger equation solutions generally demonstrated high performance, though solving quantum spin glass remained challenging. We used a simple product state $\ket{+}^{\otimes N}$ as an initial state in this bench marking, which might have limited performance in these cases. 
Interestingly, success rates on D-Wave's QA devices in solving classical problems are comparable to idealized simulation results in NISQ algorithms particulally in max-cut and knapsack problems.

\begin{figure}[H]
    \centering
    \includegraphics[keepaspectratio, width=84mm]{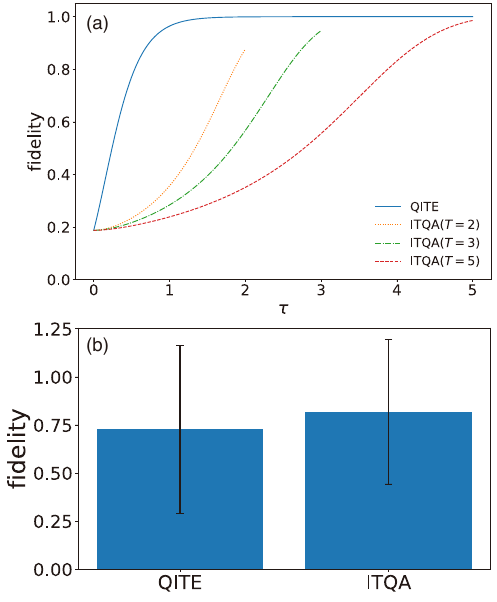}
\caption{
(a) Imaginary-time evolution of fidelity with respect to the solution state in the max-cut problem for QITE and ITQA, with various annealing times $T$ applied in ITQA. 
Results are obtained from a direct solution of the imaginary-time Schr\"odinger equation.  
(b) Fidelity in the ansatz-based QITE and ITQA for the number partitioning problem, using a longer time parameter $T = 1$ compared to Fig.~\ref{fig:1}.
}
    \label{fig:2}
\end{figure}

\begin{figure}[H]
    \centering
    \includegraphics[keepaspectratio, width=84mm]{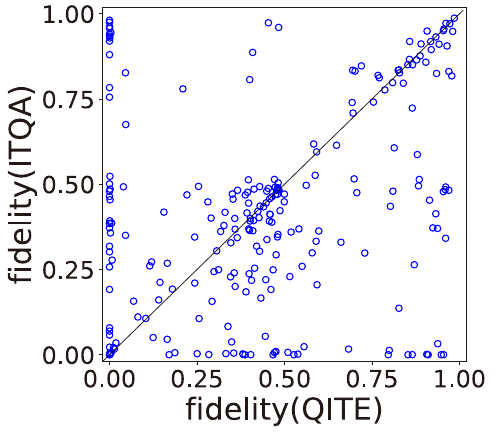}
\caption{
Fidelity scatter plots for ansatz-based QITE and ITQA in the quantum spin glass model.
}
    \label{fig:3}
\end{figure}

The ITQA harnesses the advantages of the QA and the imaginary-time evolution, enhancing ground-state fidelity; even when the state shows transition to an excited state, imaginary time evolution can re-amplify ground state populations. Nevertheless, QITE typically outperforms ITQA  (Fig.~\ref{fig:2}(a)). This is because in ITQA, the problem Hamiltonian becomes fully encoded at the end of annealing, while in QITE, it is encoded throughout imaginary-time evolution. 

Performance in ansatz-based QITE and ITQA depends on circuit structure and parameters $\Delta t$ and $T$. 
For example, at extended $T=1$, ITQA can outperform QITE on average, though variance remains high (Fig.~\ref{fig:2}(b)). 
Success probability is highly instance-dependent (Fig.~\ref{fig:3}), with neither QITE nor ITQA consistently dominating. 
Specific instances show high performance for ITQA and, in other cases, both methods excel or underperform.

\begin{figure}[t]
     \centering
     \includegraphics[keepaspectratio, width=84mm]{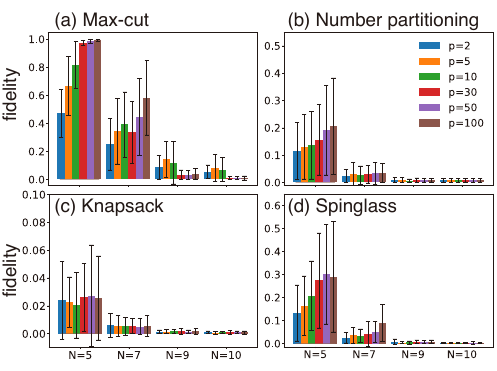}
\caption{
$p$-dependence and qubit number dependence of fidelity in QAOA across four distinct problems.
}
     \label{fig:4}
 \end{figure}

The QAOA fidelity significantly degrades for number partitioning and knapsack problems (Figs.~\ref{fig:1}(b), (c)). Larger depths $p$ are generally expected to increase QAOA success probability, though in max-cut, deeper $p$ reduces success in contrary to expectation (Fig.~\ref{fig:4}(a)). 
For knapsack, $p$-dependence becomes negligible even at small sizes, and success probability declines sharply with increasing problem size, indicating higher computational demands in QAOA.

Our benchmarking of non-FTQC algorithms provides a foundational baseline for assessing future advancements across a range of hard optimization problems. By employing default configurations and deliberately avoiding algorithm-specific improving techniques, this study offers a controlled reference for performance evaluation. Several advanced techniques, including modular parity QAOA~\cite{PRXQuantum.3.030304}, error-corrected QA~\cite{Pudenz2014}, shortcuts to adiabaticity in QA~\cite{delCampo_2019,RevModPhys.91.045001,PhysRevA.95.012309,doi:10.1073/pnas.1619826114}, and nonlocal approximations for QITE~\cite{Nishi2021}, show significant potential for improvements. Our findings also highlight the implications of the “no free lunch” theorem in optimization~\cite{585893}, which suggests that no single algorithm among non-FTQC approaches consistently outperforms others across all problem types. Consequently, tailored algorithms and alternative algorithmic strategies are critical for achieving optimal solutions across various optimization landscapes. 
Further research is required to develop more comprehensive benchmarking standards 
--- such as incorporating metrics like time-to-solution --- that will allow for a more complete and rigorous evaluation.

\section{Conclusion}

In this study, we conducted a comprehensive benchmarking of various non-fault-tolerant quantum computing (non-FTQC) algorithms across a range of combinatorial optimization problems, including max-cut, number partitioning, and knapsack, as well as quantum spin glass. 
The benchmark encompassed a diverse set of noisy intermediate-scale quantum (NISQ) algorithms --- such as the variational quantum eigensolver, quantum approximate optimization algorithm, quantum imaginary time evolution, and imaginary time quantum annealing --- as well as classical simulated annealing and quantum annealing on D-Wave devices. 
Our findings demonstrate that, under standardized configurations, no single non-FTQC approach consistently outperforms others across all problem types. 
This outcome highlights the inherent complexity and diverse requirements of combinatorial optimization tasks. 
This work establishes an essential performance baseline, underscoring the importance of adaptive strategies in advancing NISQ algorithms and quantum annealing platforms.

\section*{Acknowledgement}
S.W. was supported by JST, PRESTO Grant Number JPMJPR211A, Japan. 

% \appendix
\bibliography{ref}% Produces the bibliography via BibTeX.

\end{document}